Research article

# Folding molecular dynamics simulation of T-peptide, a HIV viral entry inhibitor : Structure, dynamics, and comparison with the experimental data.

Ioanna Gkogka & Nicholas M. Glykos*

*Department of Molecular Biology and Genetics, Democritus University of Thrace, University campus, 68100 Alexandroupolis, Greece, Tel +30-25510-30620, Fax +30-25510-30620, https://utopia.duth.gr/glykos/ , glykos@mbg.duth.gr*

# Abstract

Peptide T is a synthetic octapeptide fragment, which corresponds to the region 185-192 of the gp120 HIV coat protein and functions as a viral entry inhibitor. In this work, a folding molecular dynamics simulation of peptide T in a membrane-mimicking (DMSO) solution was performed with the aim of characterizing the peptide's structural and dynamical properties. We show that peptide T is highly flexible and dynamic. The main structural characteristics observed were rapidly interconverting short helical stretches and turns, with a notable preference for the formation of $\beta$-turns. The simulation also indicated that the C-terminal part appears to be more stable than the rest of the peptide, with the most preferred conformation for residues 5-8 being a $\beta$-turn. In order to validate the accuracy of the simulations, we compared our results with the experimental NMR data obtained for the T-peptide in the same solvent. In agreement with the simulation, the NMR data indicated the presence of a preferred structure in solution that was consistent with a $\beta$-turn comprising the four C-terminal residues. An additional comparison between the experimental and simulation-derived chemical shifts also showed a reasonable agreement between experiment and simulation, further validating the simulation-derived structural characterization of the T-peptide.

# 1. Introduction

The exterior envelope glycoprotein of HIV, gp120, plays a significant role in receptor binding and interactions with neutralizing antibodies. Structural information regarding gp120 is essential for the determination of the mechanism of HIV infection and the design of new therapeutic approaches[1]. It has been suggested that the interaction between HIV-1 and its host cell receptors could entail the region 185-192 of the gp120 coat protein[2], which corresponds to the gp120 V2 region[3]. The synthetic octapeptide fragment with the sequence: ASTTTNYT, is known as peptide T due to its high threonine content, and it was proven to function as a viral entry inhibitor by blocking the binding of both isolated gp120 and HIV-1 with the CD4 receptor[2-4]. Later studies have suggested that both the CD4 receptor and a co-receptor are needed for the invasion of healthy cells by HIV-1[2].

The folding behavior of peptide T has been studied by Picone and her colleagues[2] by means of NMR spectroscopy at 500 MHz. More specifically, NMR spectra were obtained at 500 MHz, double-quantum-filtered (DFQ) COSY, and NOESY spectra were run, and chemical shifts of all backbone protons and temperature coefficients of the labile protons were reported. The chemical shift data indicated a non-random conformational state. It was shown that residues S2 and T8, whose resonances of the NH groups were broader than the other five, could possibly adopt a single preferred conformation and that the side chains of two of the four threonines, whose methyl groups both resonate at 1.03 ppm, were in a similar environment. The same study,[2] also identified in the NOESY spectrum effects between chemical groups belonging to the four C-terminal residues, a finding that was also consistent with the presence of well-defined conformers. They concluded that peptide T demonstrates an unusual degree of conformational order in the given (DMSO) solvent environment . The minimal value of the T8 chemical shift in the range of 298-330K, the diagnostic NOE between the NH groups of Y7 and T8, and variable temperature data were interpreted as being consistent with a type I β-turn involving the four C-terminal residues, although the possibility of a helical conformation could not be positively excluded. Although this conformation was proposed to be a prominent conformation in solution, it was clear that other peptide conformations were also present, but their structural characterization was limited by the non-linear dependence of the NOEs on interatomic distances[5].

Here, we attempt to characterize the structure and dynamics of the T-peptide in DMSO using a 4.0μs-long folding molecular dynamics simulation which was performed in explicit solvent and with full treatment of the electrostatics. We extensively characterize the structural properties of peptide T, and examine which are the most likely conformations that it can adopt in this membrane-mimicking solvent. Due to the availability of the experimental chemical shift data, we are also able to critically evaluate the agreement between the simulation-derived and experimentally determined values with the aim of validating and strengthening the conclusions derived from the simulation.

In the following sections, we describe the simulation protocol, the trajectory analysis, and the results obtained from a structural analysis of the trajectory. Since one of the major concerns when evaluating the effectiveness of MD simulations of proteins or peptides is the degree to which the simulations faithfully sample the conformational space of the protein or peptide, we also examine the degree to which the simulation sufficiently samples the folding landscape of peptide T.

## 2. Methods

**2.1 System preparation and simulation protocol**

The system preparation process and the simulation protocol have been previously described in detail.[6-12] In summary, the procedure is as follows. The starting peptide structure in the fully extended state, along with the addition of missing hydrogen and the solvation-ionization of the system, were all performed using the program LEAP from the AMBER tools distribution[13]. The simulation was conducted using periodic boundary conditions with a cubic unit cell sufficiently large to guarantee a minimum separation between the neighboring cells of at least 16 Å. We studied the dynamics of the folding simulation of peptide T using the program NAMD[14,15] for a grand total of 4.04 μs using DMSO as solvent, the AMBER99SB-STAR-ILDN force field[16-18], and the adaptive tempering method[19], as implemented in the program NAMD (adaptive tempering is formally equivalent to a single-copy replica exchange folding simulation with a continuous temperature range). For our simulation, the temperature ranged between 280 and 380K inclusive. This temperature range was applied to the system through the Langevin thermostat as described below.

The simulation protocol was the following: the system was first energy minimized for 2000 conjugate gradient steps and then the temperature was increased with a ΔT step of 20K until the final desired temperature of 320K over a period of 32 ps. Subsequently, the system was equilibrated for 10 ps under constant temperature and pressure (NpT conditions) until the volume equilibrated. Then the production NpT run followed, with the temperature and pressure controlled using the Nosé-Hoover Langevin dynamics and Langevin piston barostat control methods, as implemented by the NAMD program, with adaptive tempering applied through the Langevin thermostat, while the pressure was maintained at 1 atm. The Langevin damping coefficient was set to 1 ps$^{-1}$, and the piston's oscillation period to 400 fs, with a decay time of 200 fs. The production phase was performed with the impulse of the Verlet-I multiple-step integration algorithm[20] was used as implemented by NAMD. The inner timestep was 2.5 fs, with nonbonded interactions being calculated every one step. The long-range electrostatic interactions were calculated every two timesteps, using the Particle Mesh Ewald method (PME)[21] with a grid spacing of approximately 1 Å and a tolerance of $10^{-6}$. The cutoff for the Van der Waals interactions was set at 8 Å, through a switching function, and the SHAKE algorithm[22] with a tolerance of $10^{-8}$ was used to restrict all bonds involving hydrogen atoms. The trajectory was obtained by saving the atomic coordinates every 1.0 ps.

**2.2 Trajectory analysis**

The programs CARMA[23] and its GUI program GRCARMA[24], along with custom scripts have been used for the majority of our analyses, including the removal of overall rotations/translations, calculation of $\varphi$, $\psi$ dihedral angles, calculation of RMSD's from a chosen reference structure, dihedral space principal component analysis (dPCA)[25-27] and corresponding cluster analysis, calculation of average structures and production of PDB files from the trajectories. Secondary structure assignments were calculated using the program STRIDE[28]. Other structural analyses were performed using the PROMOTIF[29] program. Chemical shifts were calculated using the program SPARTA+[30]. In order to be able to evaluate our results and make quantitative comparisons between the experimentally determined and the simulation-derived chemical shifts, we used two statistical analyses: the reduced $x^2$ and the linear correlation coefficient. All molecular graphics representations and figure preparation were performed with the programs VMD[31] and CARMA.

# 3. Results

**3.1 Extent of sampling**

Peptide T –as will be discussed in the following paragraphs– is highly flexible. This highly dynamic behavior is associated with a rugged folding energy landscape and the absence of a well-defined gradient towards a putative 'native' conformation, suggesting that a folding molecular dynamics simulation must sample the vast conformational space associated with the disordered state. To ascertain whether our simulation was adequately sampled we apply a recently proposed probabilistic method for estimating the convergence of molecular dynamics trajectories. This method is based on the application of Good-Turing statistics to estimate the probability of observing new (previously unobserved) peptide conformations. This probability is expressed as a function of the RMSD (of these new conformers) from the structures that have already been observed in the simulation[32]. The results from these calculations are shown in Figure 1.

In these diagrams (Figure 1), the high probability values of unobserved conformations for low RMSD values show –as expected– that it is highly probable to observe structures that are very similar with some of the already observed structures, with only minor differences from them. As the RMSD values from the already observed structures increase, the corresponding probability values decrease. Therefore, the exact form of these graphs as well as the rate by which they approach low probability values can inform us about the significance of the structural variability that has been missed due to limited sampling, or equivalently, the extent to which the trajectory has been adequately sampled.

The results obtained from the application of Good-Turing statistics to our trajectory are presented in the form of three independent calculations. In the first calculation we considered the entire octapeptide (black upper curve in Figure 1), in the second we excluded the two hypermobile N- and C-terminal residues (orange curve) , and for the last calculation we only used residues 5-8 to obtain the Good-Turing estimates (blue curve in Figure 1). At low RMSDs, all three curves have high probability values and for gradually increasing RMSDs, the curves asymptotically approach low probability values. The results from the analysis performed using all residues (black curve) suggest the most different structure we should expect to observe if we doubled the simulation time would differ by no more than approximately 2.0 ± 0.1 Ångström (RMSD) from those already observed. To give a worked example of the information contained in these graphs, if the simulation were to continue, we would expect that on average only one out of 25 previously unobserved structures ($P_{unobserved}$ = 0.04) would differ by an RMSD value of 1.6Å (or more) from the already observed structures. The results from the analysis performed using the residue selection consisting of residues 2-7 (orange curve in Figure 1) suggest that the most different structure we should expect to observe if we doubled the simulation time would differ by no more than approximately 1.3 ± 0.1 Ångström (RMSD) from those already observed. Finally, when using only residues 5-8 (blue curve in Figure 1) we find that the most different structure we should expect to observe if we doubled the simulation time would differ by no more than approximately 0.4 ± 0.08 Ångström (RMSD) from those already observed. Clearly, the effect of limiting the residue selection is rather dramatic: the curve quickly falls to small probability values of unobserved species for RMSD values of lower than 1Å. It can be seen that by limiting the residue selection to the amino acids comprising the C-terminal part of the peptide, the curve approaches low probability values faster, demonstrating that residues 5-8 correspond to more stable peptide conformers with a greater tendency to promote the formation of secondary structure patterns.

In summary, the conclusions derived from the application of Good-Turing statistics indicate that the length of the simulation is sufficient for allowing a reasonable sampling of the dynamics and the structural variability of the peptide for the given force field. This is especially true for the C-terminal part of peptide T which appears to adopt a much more stable behavior than the rest of the peptide.

**3.2 T-peptide is highly flexible with a pronounced preference for forming *β*-turns**

Figure 2 shows the results obtained from the secondary structure analysis of the T-peptide. Figure 2A depicts the per-residue secondary structure assignments of the peptide as a function of simulation time. It is important to mention that for this analysis we have used the entire trajectory since no considerable differences were observed when using only the structures that correspond to stable peptide conformers. As it can be seen in Figure 2A, peptide T is highly flexible and the majority of residues are being assigned to turn (cyan) or coil (white) states, while assignments to helical structures (α-helices and $3_{10}$-helices) are significantly less frequent. To better characterize the peptide's secondary structure preferences, we calculated two WebLogo diagrams[33]. The first diagram (Figure 2B) is a representation of the per residue STRIDE-derived secondary structure assignments obtained by analyzing all peptide structures observed in the simulation. For the second diagram (Figure 2C) only frames with an adaptive tempering temperature of less than 320 K were used in an attempt to increase the weight of the more stable (from the simulation's point of view) peptide conformations. The isolation of stable conformers as a function of temperature was possible because the simulation was performed using the adaptive tempering method which automatically adjusts the thermostat according to the energy of the system. A closer examination of these WebLogo diagrams reveals that there are no pronounced differences between the two calculations, and, additionally, clearly indicate that the termini are quite flexible. This behavior is expected for such a short peptide. Residues 3-5 tend to form mostly turns, while we can identify some minor occurrences of coil, $3_{10}$-helical, and even α-helical structures.

The experimental data of Picone and her colleagues[2] suggested that peptide T could adopt fairly stable conformations and proposed that a helical segment (either α-helical or $3_{10}$-helical) could be present, but the most tenable hypothesis was the one of a *β*-turn[5] forming between the C-terminal residues 5 to 8 (inclusive). These are the main structural characteristics observed in our analysis as well, but the difference between those two analyses is that our results suggest a significant degree of flexibility in the system.

In order to further analyze the main structural characteristics (turns and helices) of peptide T and have a more comprehensive understanding of the preferences among the different β-turn and helical motifs, we used the PROMOTIF program. Firstly, we analyzed the β-turns. We studied the presence and the corresponding frequency of each β-turn type, using the whole trajectory. Table I shows the frequency of each β-turn class for each of the five possible tetrapeptides contained in T-peptide, namely : a. 1-Ala-2-Ser-3-Thr-4-Thr, b. 2-Ser-3-Thr-4-Thr-5-Thr, c. 3-Thr-4-Thr-5-Thr-6-Asn, d. 4-Thr-5-Thr-6-Asn-7-Tyr, and finally, e. 5-Thr-6-Asn-7-Tyr-8-Thr. Of the turns located, β-turns type I and IV appear to be the most prominent turn motifs, while β-turn types II, VIII, I' and II' are significantly less frequent. According to the above results, the most prominent β-turn type for the amino acid sequences: 1-Ala-2-Ser-3-Thr-4-Thr and 2-Ser-3-Thr-4-Thr-5-Thr appears to be a type I β-turn, while the second most preferred β-turn type is type IV. For sequences 4- Thr-5-Thr-6-Asn-7-Tyr and 5-Thr-6-Asn-7-Tyr-8-Thr, the most preferred β-turn type is type IV, followed by a type I β-turn. It should be noted, however, that the type IV turns are not defined by a specific geometry, and only serve as a catch-all term for a β-turn not belonging to any of the other characterized types.

The next step was to analyze the second most prominent, according to our calculations, secondary structure element, helices. Table II shows the frequency of each helix type for every possible combination of sequential amino acid residues. According to the results presented in Table II, of the helices located, $3_{10}$-helices appear to be the most preferred helical motif, followed by α-helices, while π-helices were only observed in a tiny fraction of the observed structures. A comparison between the results obtained from tables I and II clearly shows a preference for β-turns rather than helices. These observation are in agreement with our previous results obtained from the secondary structure analysis and shown in Figure 2. Also, it is important to mention that according to our calculations, the most preferred conformation for the amino acid sequence 5-Thr-6-Asn-7-Tyr-8-Thr is a β-turn type IV (5.72%), followed by a β-turn type I (4.41%), while the percentage of the helical conformations at the C-terminal end of the peptide range in the vicinity of values of about ~2-3% (the absence of helical assignments involving the first and last residues is a consequence of the algorithm used in PROMOTIF and should not be taken at face value). To summarize, our observations based on secondary structure analysis is in good agreement with the experimental conclusions, and indicate that the most prominent conformation is a 5-8 *β*-turn rather than a C-terminal helical segment[5]. The fact that the experimental results indicated that the most likely cyclic structure is a type I *β*-turn[5] is also in agreement with our results if we take into account that the type IV are in reality an heterogeneous collection of turns not assigned to other types.

**3.3 Dihedral principal component analysis allows visualization of the peptide's structural characteristics.**

The secondary structure analysis described above enabled us to recognize the basic structural characteristics of the peptide. In this section, and in order to place our observations in a more structurally oriented framework, we analyze the folding landscape of the peptide using dihedral angle Principal Component Analysis (dPCA), which also allows us to identify the most prominent peptide conformations. Previous studies have shown that dPCA is an attractive method because the analysis starts with the relevant part of the dynamics, preventing unnecessary noise. Moreover, and since dPCA is based on the backbone dihedral angles, it can easily distinguish between the kinetically well separated main conformational states of the peptide, such as the $\alpha_R$ helical and the *β* extended conformations[25-27]. As described in section 2.2 dPCA was performed and an initial set of clusters was identified using the programs CARMA and GRCARMA. The results from these calculations will be presented in the form of two-dimensional (log density) projections of the trajectory along its top three principal components as shown in Figure 3. In these diagrams, high-density peaks are illustrated as dark blue regions and correspond to clusters of structures with similar principal component values, and as a result similar dihedral angles and backbone structures. As shown in Figure 3, we performed two sets of calculations. In the first set (upper row of graphs in Figure 3), all peptide residues were used for the dPCA analysis. The resulting landscapes are blurry and complex, faithfully illustrating the structural complexity of this highly flexible peptide. Limiting, however, the selection of residues used for dPCA to the C-terminal half of the peptide (residues 5-8, lower row of diagrams in Figure 3) gives a completely different picture : there is a small number of distinct and well-defined minima corresponding to structurally and kinetically stable peptide conformers which are continuously interconverting between them. The limited number of distinct structures suggests that the C-terminal segment of the peptide adopts a limited number of stable conformations corresponding to distinct secondary structures with specific torsion angles and hydrogen bond patterns. These results are in agreement with the experimental conclusion that residues 5-8 of peptide T adopt a more stable conformation.

Having acquired the distribution of the principal components from the dPCA analysis, it is feasible to relate high-density peaks with distinct peptide conformations (note that the actual cluster analysis was performed in the three-dimensional space defined by the top three principal components, and not in two dimensions as shown in Figure 3 for clarity). Representative structures for the clusters were identified by calculating an average structure for each cluster and then selecting the frame from the trajectory with the lowest RMSD from the corresponding average structure. To bring forward the structural heterogeneity present in these clusters, in Figure 4 we show not just a single structure (the representative), but a whole set of peptide structures that belong to these clusters. In the center of this figure is the log density projection of the trajectory on the first two principal components derived from dPCA. The marked peaks (1 to 7) are in a one-to-one correspondence with the structural diagrams at the periphery of the diagram. The numbers below each schematic structure representation are the relative percentages of simulation time that each of the clusters occupied. We should mention here that CARMA will only assign a structure to a cluster if the structure is very near the core region of a peak in the dPCA map. The result is that not all structures from the trajectory are assigned to clusters. This can be seen in Table III which shows that in only 2,894,280 structures (corresponding to 57.24% of the total number of 5,056,000 structures recorded) were assigned to clusters. The three main clusters recorded for peptide T were cluster 1, cluster 2, and cluster 4 which occupied approximately 16%, 12%, and 9% of the whole trajectory, or, equivalently, 28%, 22% and 16% of the total number of assigned peptide structures. These relatively low percentages clearly indicate again the flexibility and dynamic behavior of this peptide.

Figure 4 clearly demonstrates that the main structural characteristics observed were turns and helices, as has already been discussed in the previous sections of this communication. Due to the increased kinetic frustration of the system, the representative structures differ between the clusters, and there is also a considerable presence of unfolded random coil representatives. To have a more detailed view of the structural variability of the system, Table IV presents the per residue STRIDE-derived secondary structure assignments of the five most populated motifs for each cluster.

The structures in Figure 4 clearly indicate the plasticity of the peptide, but –by being the structures closest to the average– they obscure the real extend of the structural variability, especially for the N-terminal residues (which were not part of the dPCA-based clustering). To bring this forward, we show in the Supporting Information Figure SI the same collection of structures, but this time using for each cluster a superposition of 500 uniformly selected structures that belong to the respective clusters. Next to each superposition figure, we have also added a WebLogo diagram specific for the cluster under examination. Naturally, the superpositions shown in this SI Figure are complex and noisy, making it difficult to understand the structural content that is present in the clusters. Nevertheless, it is clear that as expected, the C-terminus forms more compact structures than the N-terminus.

To summarize this section, T-peptide shows a preference for mostly extended structures, rich in transiently stable turns and, less frequently, helical structures. These structures are quickly and constantly interconverting as Figure 2(A) clearly indicates. Characterization of the folding landscape using dPCA, showed that the C-terminal residues appear to be better behaving structurally, and allowed the generation of meaningful structure schematics such as those shown in Figure 4. Based on this analysis, it is clear that any attempt to define a single stable "native" structure for the T-peptide is meaningless, and this we believe is the major difference with the experimental work of Picone *et al*[5] : what the dynamics show is that the peptide should be characterized in terms of average statistical preferences, and not in terms of a single, kinetically stable "native" structure. It is, thus, clear that a direct comparison with NMR deliverables, such as chemical shifts, offer a much better ground for a comparison with the actual experiment than the one offered by comparing assumed structural characteristics of such a flexible and dynamic peptide.

### 3.4 The experimental and simulation-derived chemical shifts are highly correlated.

In the two previous sections, we investigated the major structural characteristics of the simulation, examined the presence and the corresponding frequency of the various *β*-turn types and helical conformations, and identified the preferred peptide conformers for a set of clusters derived from the corresponding dPCA analysis. In this section, we aim to make a more detailed and quantitative comparison between the experimental and simulation-derived results. The NMR experiments of Picone *et al*. provided the numerical values of the H-chemical shifts of peptide T in DMSO[5]. Unfortunately, neither the NMR data nor any structural models have been deposited in either the PDB or BMRB databases, nor could these data been obtained from the authors. We have, however, been able to obtain the NMR-derived chemical shifts directly from the printed form of that paper and to compare the simulation-derived chemical shifts with their corresponding experimental values.

Before presenting the procedure we have followed for the calculation of the simulation-derived chemical shifts and the results obtained from the direct numerical evaluation of the differences between the simulation and the experiment, it is important to highlight the effect of solvent on the NMR chemical shifts. The dependence of H chemical shifts on solvent has been thoroughly studied since the beginning of high-resolution proton NMR[34]. Buckingham et al.[35] described four interactions responsible for solvent effects, namely: hydrogen bonding, the anisotropy of the solvent molecules, polar and van der Waals effects, the impact of each one of these contributions can vary significantly. Since it was not possible to include the effect of the organic solvent on the calculation of the simulation-derived chemical shifts, it is important to note that due to these constraints, our calculations may present certain limitations as will be discussed below.

The process followed for our calculations was the following: intending to obtain the experimental secondary chemical shifts, we used the experimental chemical shifts from the Picone *et al*. publication[5] and the 1H random coil chemical shifts for peptides of sequence GGXAGG (where X is any of the 20 naturally occurring amino acids or the modified amino acid 4-hydroxyproline) measured in DMSO from Tremblay, Banks & Rainey[36]. For the calculation of the simulation-derived secondary chemical shifts, we first calculated the simulation-derived chemical shifts using the programs CARMA and SPARTA+ through a perl script, taking into account all 5056000 frames of our trajectory and the random coil chemical shifts from the Tremblay, Banks & Rainey publication[36]. This perl script produces pdb files for each frame of the trajectory, the pdb files are read by SPARTA+ and then the chemical shifts, as well as the mean and the standard deviation values are calculated. The detailed numerical results are shown in Tables SI and SII of the Supplementary Information file, but before discussing in detail those results, we must evaluate the limitations of the methodology we applied for the comparison between the experimental and simulation-derived shifts.

As already hinted above, the basic problem with this analysis is that, due to lack of suitable analytical tools, it is not feasible to calculate chemical shifts in DMSO for the approximately five million peptide structures obtained from the simulation. Tremblay, Banks & Rainey[36], however, have shown that the secondary chemical shifts are more affected by protein secondary structure than solvent environment. We perceived this as an opportunity for proceeding with the calculation as follows. In the first step, expected chemical shifts of the peptide structures *in water* are calculated using established procedures, in our case, the SPARTA+ program. Because of the effect of solvent (DMSO vs water), these calculated shifts will differ significantly from the experimental, but this difference will be mostly a difference in scale and not a difference in the exact motif that the shifts follow along the peptide chain. Note that the expectation that the differences between the measurements can be approximated as a residue-dependent shift of the values assumes (a) that the chemical shifts are indeed more affected by protein secondary structure than solvent environment, and, more importantly, (b) that the simulation correctly captures the structural ensembles visited by the T-peptide. If both of these assumptions hold true, then we would expect that the raw experiment vs simulation shifts would be highly correlated but with a systematic translation/drift of the values arising from the different shielding or deshielding effect that DMSO has on specific atomic types. Figure 5 shows a direct comparison between the HA and HN chemical shifts as obtained from experiment and simulation. For the HA shifts, the graph shows the unscaled (raw) data, for the HN case the two graphs have been scaled to one another. What Figure 5 clearly shows is that the experimental and simulation-derived shifts are indeed highly correlated (with respective correlation coefficients of 0.78 and 0.87 for HA and HN), but, as expected, translated relative to one another due to the effect of the solvent.

The findings of Figure 5 do indicate that the simulation meaningfully captures the structures and dynamics of the peptide in DMSO, but due to the offset of the chemical shift values, we can not properly quantify this agreement in terms of a suitable statistic like the reduced $\chi^2$. We attempted to correct for this problem of scale by applying an additional correction to the simulation-derived shifts as follows. Our argument rests on the assumption that a useful numerical estimate for the shielding/deshielding effect of DMSO (relative to water) can be obtained by comparing the differences between the *random coil shifts* for the various amino acids in DMSO and water. Values for the random coil shifts in these solvents (and for the various amino acids) are readily available, and these differences can be applied to the simulation-derived chemical shifts to bring them on approximately the same scale as the experimental values. This procedure may sound reasonable, but introduces additional errors that are hard to estimate, with the most important being that the random coil shifts are not some well-known fixed values, and various estimates of these can differ even by a factor of 2 (see Figure 2 of Tremblay, Banks & Rainey[36] and references therein for a clear indication of the amount of deviation observed when using different estimates of the random coil shifts). With these precautions, we show in the Supplementary Information Table SI the complete listing of both the experimental and simulation-derived shifts (including the correction for the random coil shifts). In this same table, we also include the estimated random coil shifts in water and in DMSO, as well as two estimated sources of error : the standard deviation obtained from the simulation *per se*, and the estimated standard error as reported by SPARTA+ for the calculation of chemical shifts in water. These two estimated errors are not independent in the sense that the detailed structural variance giving rise to the variance of chemical shifts is already incorporated in the SPARTA+ standard deviations. Given that the SPARTA-derived standard deviation is almost always higher than the simulation-derived, we resorted to using only the SPARTA-derived values as the sole source of estimated error of the simulation-derived chemical shifts.

Having obtained the primary data shown in Table SI, we can now calculate the experimental and simulation-derived HA and HN secondary chemical shifts as shown in the Supplementary

Information file Table SII. For this calculation we have used the corrected simulation-derived data and the random coil shifts of Tremblay, Banks & Rainey[36]. The reasonable agreement between the experimental and simulation-derived secondary shifts is obvious even from the raw data, and it is made clear in Table V of the manuscript in which we tabulate the values of the reduced $\chi^2$ and linear correlation coefficient statistics for various combinations of the derived quantities.

The values of the linear correlation coefficients in Table V are all in the ~0.80 range, reaching values of even ~0.87 for the secondary shifts of the HA atoms. Such high values clearly indicate that the simulation correctly captures the kinetics and the structural ensembles visited by the T-peptide. We must add here that this agreement between the NMR data and the simulation is obvious even at the level of raw data (without our correction for random coil shifts) as we already showed in the diagrams of Figure 5. Turning our attention to the reduced $\chi^2$ values, we note that all of them are lower than 1.0, reaching values as low as ~0.25 for the HN chemical shifts. We believe that such low values of $\chi^2$ are not so much the result of an excellent agreement between experiment and simulation, but that they are a by-product of the possibly overestimated standard deviation values reported by SPARTA+. One last observation concerns the fact that the residues with the largest deviations between experiment and simulation are two terminal residues, Ala1 and Thr8. This observation may indicate that the termini are not as disordered as the simulation indicated, and that, for example, the turn-forming propensity of the peptide's C-terminus may by significantly higher than the approximately ~10% frequency estimated from the simulation (Table I). Whether this is indeed the case, and whether this points to a deficiency in the force field parameters of the peptide, or of the DMSO, it is impossible to ascertain from the presently available data.

To summarize this section, we have shown a reasonable agreement between the experimental and simulation-derived chemical shifts for the T-peptide. The simulation, even without any additional correction for the solvent differences, correctly captures the variation of chemical shifts along the peptide chain (Figure 5). By using the established differences between the random coil values for the two solvents (DMSO and water), we attempted to empirically correct for the solvent effects on chemical shifts and obtained a significantly improved agreement between experiment and simulation, with values of the linear correlation coefficient of approximately 0.80, and reduced $\chi^2$ values close to 1.0. Having said that, we believe that little confidence must be placed upon our *ad hoc* attempt to obtain quantitatively useful shifts in DMSO. The shear number of assumptions and approximations involved, makes the computationally corrected shifts questionable. It is only the direct comparison of the uncorrected shifts of Figure 5 that gives us any confidence that the simulation captured the peptide's essential dynamics.

# 4. Summary and Conclusions

The prime objective of this communication was to validate the application of molecular dynamics simulations for predicting the structure and dynamics of peptide T through a comparison with the experimental (NMR) data obtained by Picone and coworkers[5]. Such an analysis is useful mainly due to the non-aqueous solvent used in both the computational and experimental work, which would validate the application of folding molecular dynamics simulations even in organic solvents.

The synthetic octapeptide fragment with the sequence ASTTTNYT is known as peptide T due to its high threonine content and it known to function as a viral entry inhibitor. Peptide T is the fragment corresponding to the region 185-192 of the gp120 HIV coat protein[2,3,4]. Picone et al. studied peptide T as a zwitterion in DMSO solution by means of NMR spectroscopy at 500 MHz. Their results suggested that a type I $\beta$-turn including the four C-terminal residues, T5, N6, Y7, and T8 was the most prominent structure. However, they also noted that this conformation was not the only one present in solution and seemed to be the only one detectable due to the non-linear dependence of NOE on interatomic distances[5].

Secondary structure analysis using the programs STRIDE and Weblogo showed that peptide T is highly flexible and that it comprises a dynamic system. The majority of residues were assigned to turn or coil states, while assignments to helical structures were very rare. Both WebLogo diagrams indicated that the first and last residues are quite flexible and correspond to coil states. Residues 3-5 tend to form mostly turns, while some minor occurrences of coil, $3_{10}$-helical, and even $\alpha$-helical structures were also identified. The above mentioned main structural characteristics were also observed by Picone and her colleagues[5], but unlike their findings, our results suggested a significant degree of flexibility in the system.

The structural analysis of turns and helices performed using the promotif program enabled us to gain a more detailed view of the specific types of turns and helices that peptide T could adopt. According to our results, the most preferred $\beta$-turn types were types I and IV, while $\beta$-turns type II, VIII, I' and II' were not so frequent. In more detail, the most prominent β-turn type for the amino acid sequences: 1-Ala-2-Ser-3-Thr-4-Thr and 2-Ser-3-Thr-4-Thr5-Thr was a type I *β*-turn, while the second most preferred *β*-turn type was type IV. For the sequences 4-Thr-5-Thr-6-Asn-7-Tyr and 5-Thr-6-Asn-7-Tyr-8-Thr, the most preferred *β*-turn type was type IV, followed by type I. Regarding the helices, the most preferred type of helix was $3_{10}$-helix, followed by *α*-helix, while *π*-helix was extremely rare. Overall, our calculations clearly showed a preference for *β*-turns rather than helices. Also, according to our calculations, the most preferred conformation for the amino acid sequence 5-Thr-6-Asn7-Tyr-8-Thr was a *β*-turn type IV, followed by a *β*-turn type I, while no helical conformations were observed for this combination and the 4-8 one. This observation is in agreement with the experimental conclusions, where it is stated that the most prominent conformation is a 5-8 *β*-turn rather than a 4-8 helical segment[5]. But unlike our findings, the experimental results state that the most likely cyclic structure is a type I *β*-turn rather than a type IV.

The dPCA analysis suggested that the 5-8 amino acid residue segment of peptide T adopts more stable conformations, which correspond to distinct secondary structures with specific torsion angles and hydrogen bond patterns. These results are in agreement with the experimental conclusions. The association of high-density peaks with distinct peptide conformers demonstrated once again that the main structural characteristics were turns and helices. Due to the increased kinetic frustration of the system, the representative structures differed between the clusters, while many coil conformations were apparent as well.

To ascertain whether our simulation was efficiently sampled we applied Good-Turing statistics. We applied this method using the Cα atoms of all residues of the peptide and then we limited the residue selection to residues 5-8. Our results clearly indicate that the structural variability of this part of the peptide has been sufficiently sampled, confirming that the C-terminal part of the peptide corresponds to more stable conformations. Lastly, the quantitative comparisons between the experimental and the simulation-derived chemical shifts showed a reasonable agreement with values of the linear correlation coefficient for the HA and HN chemical shifts were close to ~0.85.

Overall, we consider this to be a successful application of a folding molecular dynamics simulation in an organic, membrane-mimicking solvent. The simulation appears to reproduce most of the findings obtained from the experimental NMR studies, and to capture the pattern of the NMR-derived chemical shifts along the length of the peptide chain, while providing a rich view of the structural malleability of the T-peptide.

# Figure captions

**Figure 1. Extent of sampling and statistical significance.** The black upper curve is the Good-Turing estimate obtained for the full-length peptide. The orange curve shows the results obtained after exclusion of the terminal residues. The blue curve is the estimate obtained from the C-terminal half of the peptide (residues 5-8 inclusive). See section 3.1 for a detailed discussion of this figure.

**Figure 2. Secondary structure analysis**. (A) Evolution of the peptide's secondary structure versus simulation time as obtained from STRIDE. The color coding is pink for α-helices, purple for $3_{10}$-helices, cyan for turns and white for coil. (B) WebLogo representation of the per residue STRIDE-derived secondary structure assignments corresponding to all the frames of the simulation. (C) WebLogo representations of the per residue STRIDE-derived secondary structure assignments corresponding to the frames of the simulation with an adaptive tempering temperature of less than 320 K. For (B) and (C) each letter corresponds to a different secondary structure element: H for α-helices, G for $3_{10}$-helices, T for turns and C for random coil.

**Figure 3. Structural analysis : dihedral principal component analysis.** Two dimensional projections of the dihedral-PCA-derived distributions obtained from the peptide structures sampled during the simulation. Two sets of the projections are shown, the first (top row) using all residues, the second (lower row) using only the C-terminal half of the peptide. For both sets, the log density distributions along the components 1-2, 1-3 and 2-3 are shown (origin at the top left corner, first component vertical). Low density areas are dark brown/yellow, high density areas are blue/dark blue.

**Figure 4. Structural analysis : Clusters, representative structures and their frequencies.** The distribution in the center of the diagram is the log density projection of the trajectory on the first two dihedral-PCA-derived principal components obtained from using only the C-terminal half of the peptide for the analysis. For the top 7 (out of 8) prominent conformers, peptide structures and relative frequencies of the corresponding clusters are indicated. The structure schematics are superpositions of representative members of each cluster, drawn and colored

according to secondary structure content (pink for α-helices, purple for $3_{10}$-helices, cyan for turns and white for coil). In all diagrams, the peptide structures have been oriented in such a way that the C-terminus is pointing upwards. The percentages below the structures are the relative frequencies of the corresponding clusters.

**Figure 5.** Per-residue comparison between the experimental (NMR) and simulation-derived chemical shifts for the HA (upper graph) and HN (lower graph) protons. For the HN case the simulation-derived shifts have been scaled to fit the experimental range of values.

**Table I**

Table I. Populations (%) of the different *β*-turn types sampled in the MD trajectories[a].

| *β*-Turn type[b] | 1A 2S 3T 4T | 2S 3T 4T 5T | 3T 4T 5T 6N | 4T 5T 6N 7Y | 5T 6N 7Y 8T |
|---|---|---|---|---|---|
| I | 11.86% | 14.83% | 11.27% | 5.93% | 4.41% |
| II | 0.19% | 0.2% | - | 0.0001% | 0.12% |
| VIII | 1% | 0.9% | 0.51% | 0.97% | 0.68% |
| I' | 0.17% | 0.37% | - | - | 0.05% |
| II' | 0.02% | 0.02% | 0.03% | 0.005% | 0.03% |
| IV | 7.17% | 12.01% | 11.81% | 6.7% | 5.72% |

[a]Shown are the % trajectory populations of the various *β*-turn types for every possible sequential amino acid tetrad. The assignment of *β*-turns was performed using the PROMOTIF program.

[b]Types VIa1, VIa2, and VIb are excluded from our analysis since they require a Pro residue in position *i*+2.

**Table II**

Table II. Populations (%) of the different types of helical conformations sampled in the MD trajectories[c].

| Residue selection | $3_{10}$-helix | $\alpha$-helix | $\pi$ helix |
|---|---|---|---|
| 2 - 4 | 3.44% | - | - |
| 3 - 5 | 3.34% | - | - |
| 4 - 6 | 5.84% | - | - |
| 5 - 7 | 1.19% | - | - |
| 2 - 5 | 1.19% | 3.11% | - |
| 3 - 6 | 1.5% | 1.64% | - |
| 4 - 7 | 1.04% | 3.52% | - |
| 2 - 6 | 1.09% | 0.59% | 0.01% |
| 3 - 7 | 0.37% | 0.64% | 0.01% |
| 2 - 7 | 0.32% | 0.51% | 0.0002% |

[c]Shown are the % trajectory populations of the various helical conformations for every possible combination of sequential amino acid residues. The assignment of helices was performed using the PROMOTIF program.

# Table III

**Table III.** Populations of the eight clusters (in number of frames) produced by dPCA along with the percentage of clustered structures.

| Cluster | Number of structures (out of 5056000) | Percentage |
|---|---|---|
| 1 | 829017 | 16.4% |
| 2 | 627294 | 12.4% |
| 3 | 360755 | 7.1% |
| 4 | 456049 | 9.0% |
| 5 | 310782 | 6.1% |
| 6 | 144079 | 2.8% |
| 7 | 128426 | 2.5% |
| 8 | 37878 | 0.7% |

## Table IV

**Table IV.** Per residue STRIDE-derived secondary structure assignments of the five most populated motifs for each cluster.

| Cluster number | Most populated motifs | | | | |
|---|---|---|---|---|---|
| | 1st | 2nd | 3rd | 4th | 5th |
| 1 | CCC**GGG**CC | CC**TTTTT**C | **TTTTTTT**C | CCC**TTTT**C | **TTT****GGG**CC |
| 2 | CCCCCCCC | C**TTTTT**CC | CC**TTTT**CC | **TTTTTT**CC | C**GGG**CCCC |
| 3 | C**TTTTTT**CC | CCCCCCCC | CC**TTTT**CC | **TTTT**CCCC | **TTTTTT**CC |
| 4 | CCCCCCCC | **TTTTT**CCC | C**TTTT**CCC | **TTTT**CCCC | C**GGG**CCCC |
| 5 | CC**TTTTT**C | **TTTTTTT**C | CCC**GGG**CC | C**TTTTTT**C | CC**GGG**CCC |
| 6 | CCCCCCCC | **TTTTT**CCC | C**TTTT**CCC | **TTTT**CCCC | C**GGG**CCCC |
| 7 | CCCCCCCC | C**TTTT**CCC | **TTTTT**CCC | **TTTT**CCCC | CB**TT**BCCC |
| 8 | CCCCCCCC | C**TTTT**CCC | **TTTT**CCCC | **TTTTT**CCC | CC**TTTT**CC |

**Table V**

**Table V.** Reduced $\chi^2$ and correlation coefficient values between the experimental and simulation-derived data for the chemical shifts (HA and HN) and the secondary chemical shifts ($\Delta\delta$, $\Delta\delta$ HA and $\Delta\delta$ HN).

|  | HA | HN | $\Delta\delta$ | $\Delta\delta$ HA | $\Delta\delta$ HN |
|---|---|---|---|---|---|
| $\chi^2$ | 0.8724 | 0.2539 | 0.5450 | - | - |
| r | 0.8428 | 0.8514 | 0.8126 | 0.8688 | 0.7865 |

**Figure 1**

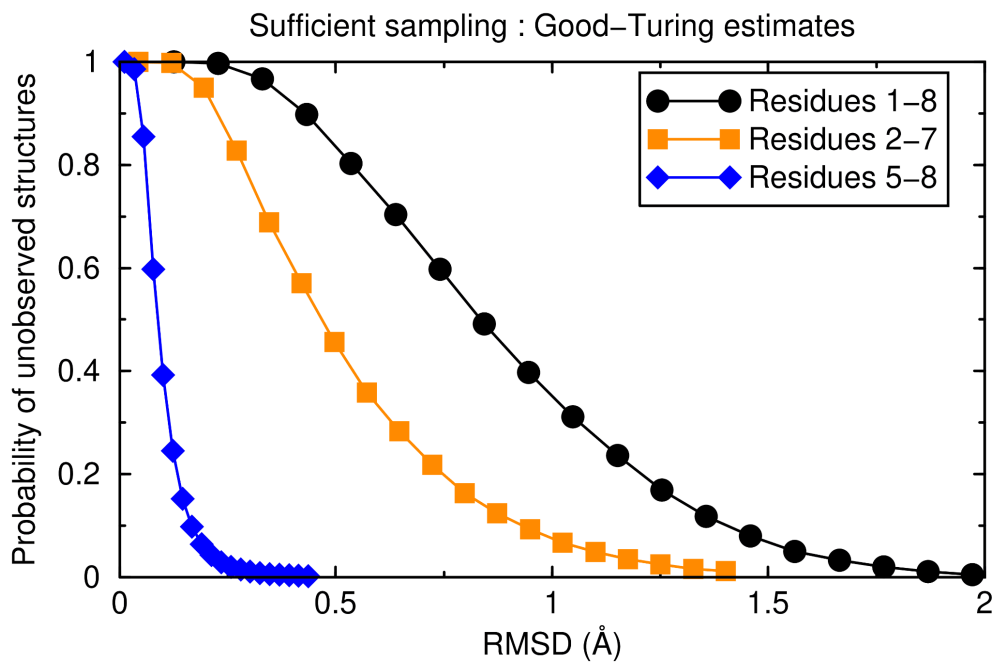

**Figure 1. Extent of sampling and statistical significance.** The black upper curve is the Good-Turing estimate obtained for the full-length peptide. The orange curve shows the results obtained after exclusion of the terminal residues. The blue curve is the estimate obtained from the C-terminal half of the peptide (residues 5-8 inclusive). See section 3.1 for a detailed discussion of this figure.

**Figure 2**

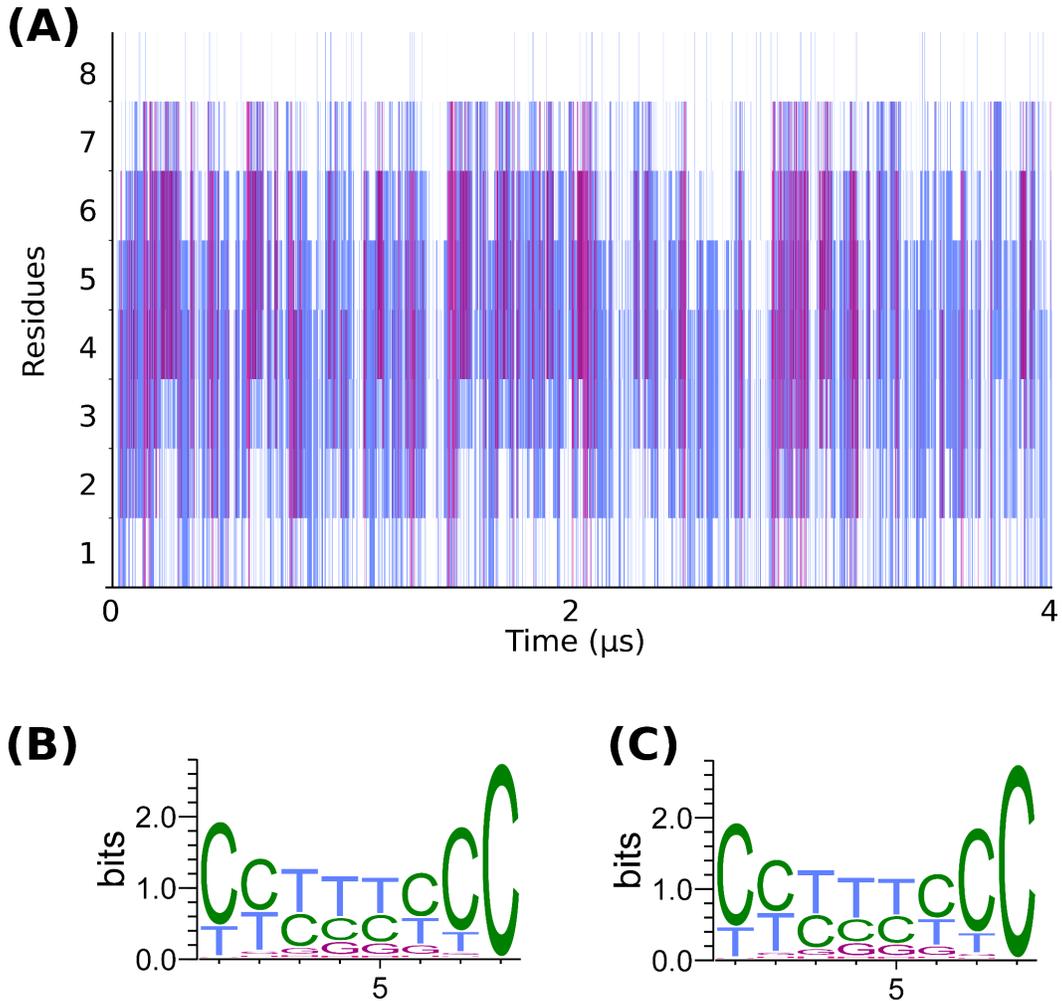

**Figure 2. Secondary structure analysis**. (A) Evolution of the peptide's secondary structure versus simulation time as obtained from STRIDE. The color coding is pink for α-helices, purple for $3_{10}$-helices, cyan for turns and white for coil. (B) WebLogo representation of the per residue STRIDE-derived secondary structure assignments corresponding to all the frames of the simulation. (C) WebLogo representations of the per residue STRIDE-derived secondary structure assignments corresponding to the frames of the simulation with an adaptive tempering temperature of less than 320 K. For (B) and (C) each letter corresponds to a different secondary structure element: H for α-helices, G for $3_{10}$-helices, T for turns and C for random coil.

**Figure 3**

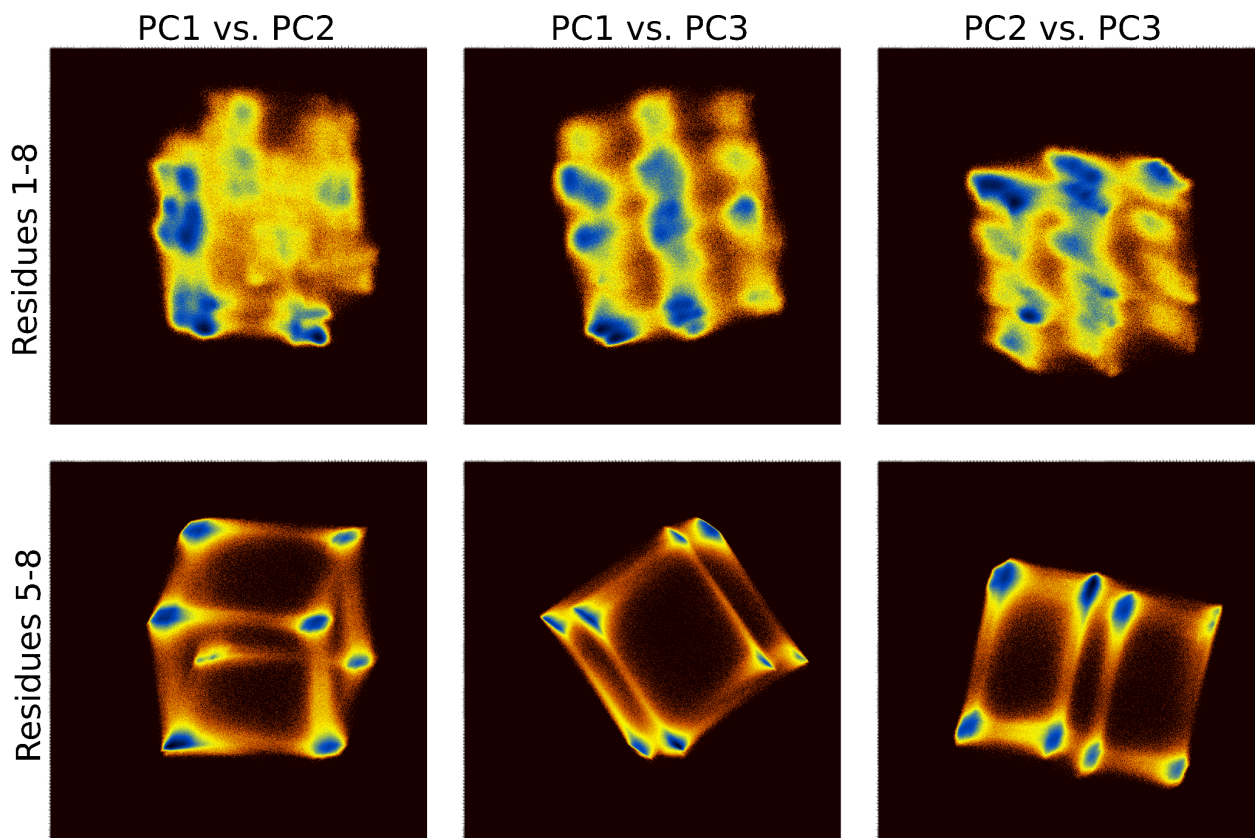

**Figure 3. Structural analysis : dihedral principal component analysis.** Two dimensional projections of the dihedral-PCA-derived distributions obtained from the peptide structures sampled during the simulation. Two sets of the projections are shown, the first (top row) using all residues, the second (lower row) using only the C-terminal half of the peptide. For both sets, the log density distributions along the components 1-2, 1-3 and 2-3 are shown (origin at the top left corner, first component vertical). Low density areas are dark brown/yellow, high density areas are blue/dark blue.

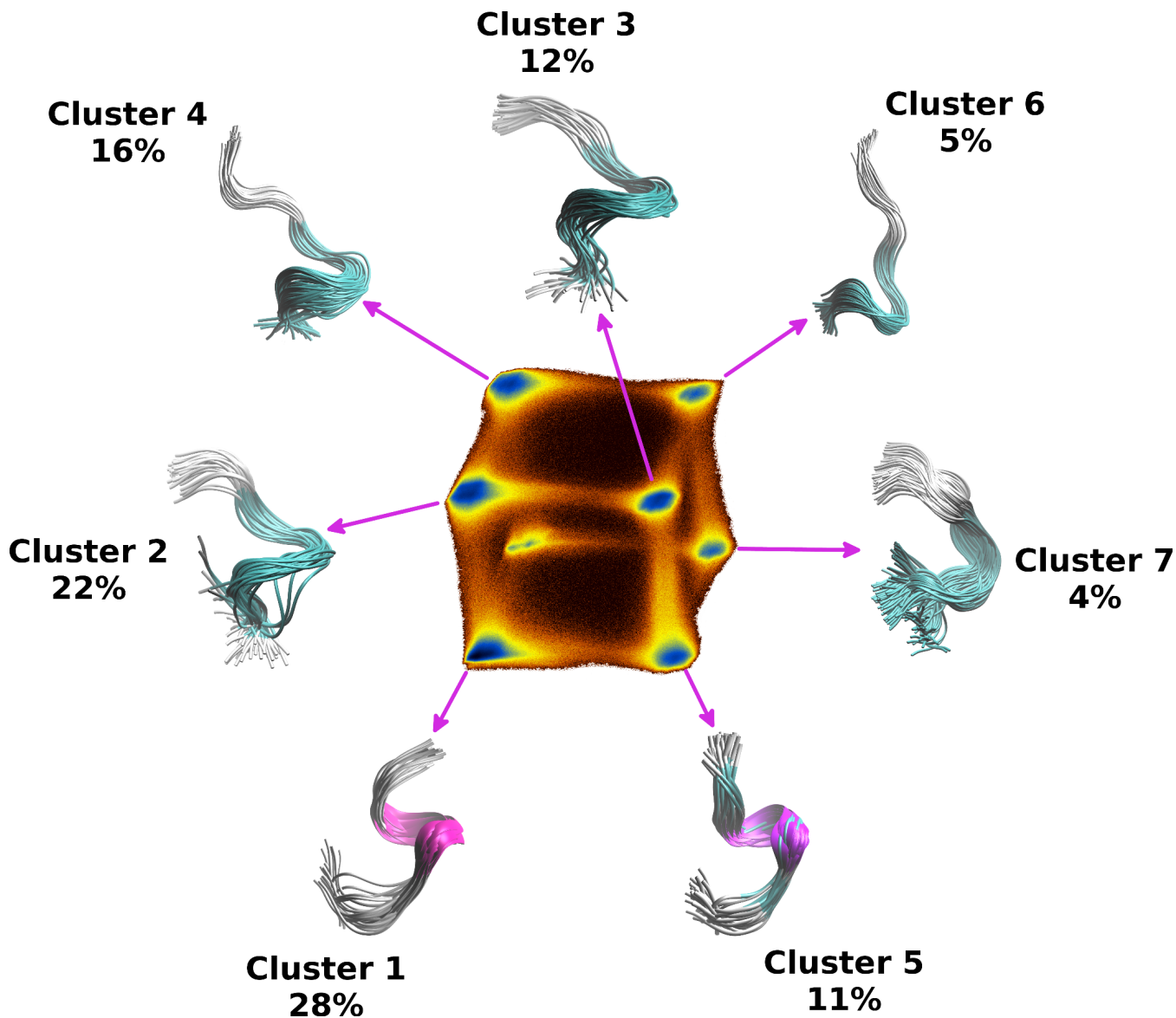

**Figure 4. Structural analysis : Clusters, representative structures and their frequencies.** The distribution in the center of the diagram is the log density projection of the trajectory on the first two dihedral-PCA-derived principal components obtained from using only the C-terminal half of the peptide for the analysis. For the top 7 (out of 8) prominent conformers, peptide

structures and relative frequencies of the corresponding clusters are indicated. The structure schematics are superpositions of representative members of each cluster, drawn and colored according to secondary structure content (pink for α-helices, purple for $3_{10}$-helices, cyan for turns and white for coil). In all diagrams, the peptide structures have been oriented in such a way that the C-terminus is pointing upwards. The percentages below the structures are the relative frequencies of the corresponding clusters.

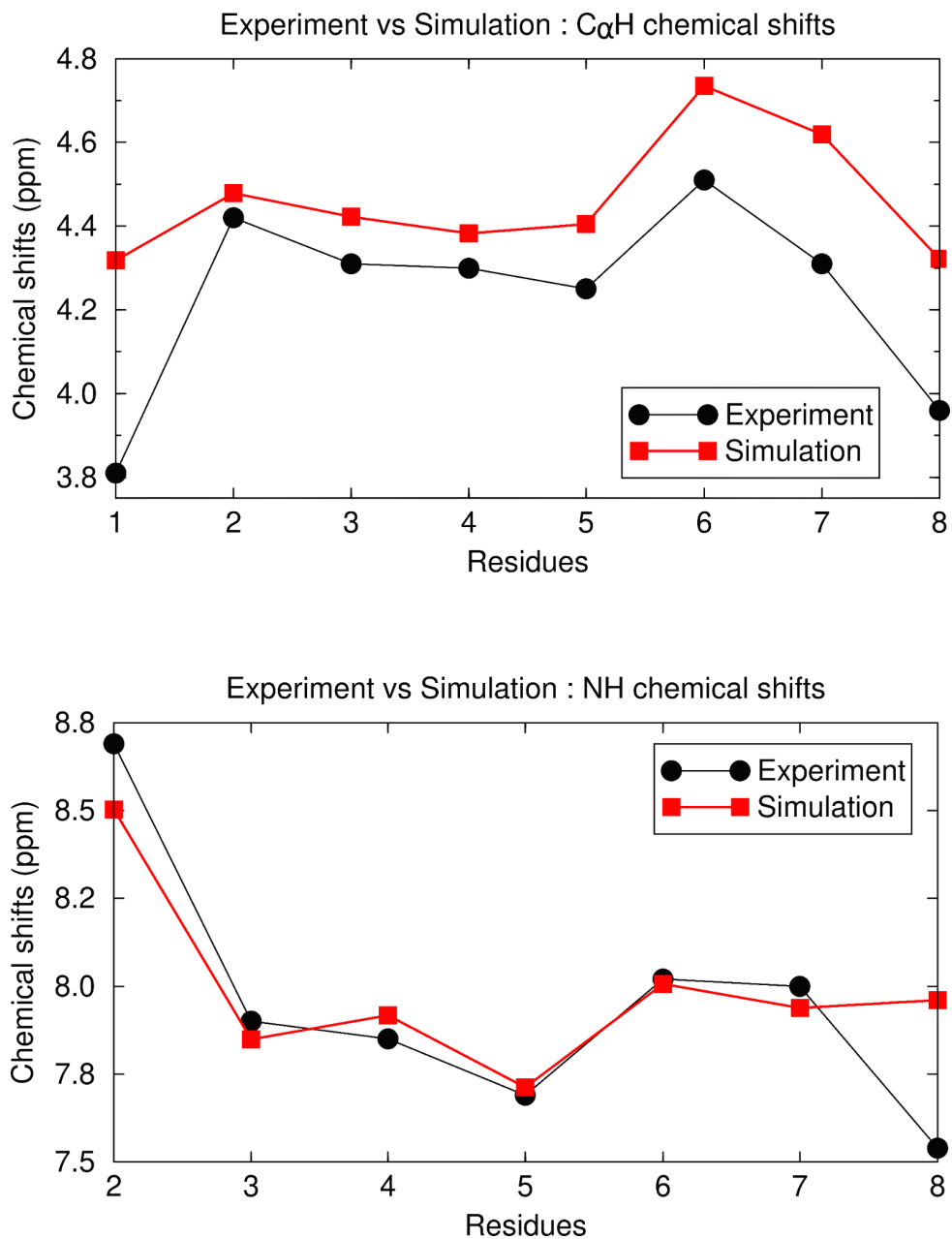

**Figure 5.** Per-residue comparison between the experimental (NMR) and simulation-derived chemical shifts for the HA (upper graph) and HN (lower graph) protons. For the HN case the simulation-derived shifts have been scaled to fit the experimental range of values.

# Folding molecular dynamics simulation of T-peptide, a HIV viral entry inhibitor : Structure, dynamics, and comparison with the experimental data.

Ioanna Gkogka & Nicholas M. Glykos*

*Department of Molecular Biology and Genetics, Democritus University of Thrace, University campus, 68100 Alexandroupolis, Greece, Tel +30-25510-30620, Fax +30-25510-30620, https://utopia.duth.gr/glykos/ , glykos@mbg.duth.gr*

# Table SI

Table SI. Experimental and simulation-derived HA and HN chemical shifts, along with the random coil and standard deviation values.

| Residue number | Residue | Atom | Experimental chemical shifts | Simulation chemical shifts Raw / Corrected | | Random coil chemical shifts in DMSO | Random coil chemical shifts (SPARTA+) | σ (simulation) | σ (SPARTA+) |
|---|---|---|---|---|---|---|---|---|---|
| 1 | A | HA | 3.81 | 4.318 | 4.168 | 4.47 | 4.320 | 0.056 | 0.217 |
| 2 | S | HA | 4.42 | 4.479 | 4.449 | 4.50 | 4.470 | 0.225 | 0.221 |
|   |   | HN | 8.69 | 8.403 | 8.463 | 8.37 | 8.310 | 0.275 | 0.531 |
| 3 | T | HA | 4.31 | 4.423 | 4.363 | 4.41 | 4.350 | 0.181 | 0.280 |
|   |   | HN | 7.90 | 8.158 | 7.918 | 8.00 | 8.240 | 0.348 | 0.531 |
| 4 | T | HA | 4.30 | 4.383 | 4.323 | 4.41 | 4.350 | 0.259 | 0.280 |
|   |   | HN | 7.85 | 8.183 | 7.943 | 8.00 | 8.240 | 0.310 | 0.531 |
| 5 | T | HA | 4.25 | 4.405 | 4.345 | 4.41 | 4.350 | 0.188 | 0.280 |
|   |   | HN | 7.69 | 8.107 | 7.867 | 8.00 | 8.240 | 0.301 | 0.531 |
| 6 | N | HA | 4.51 | 4.735 | 4.735 | 4.74 | 4.740 | 0.183 | 0.186 |
|   |   | HN | 8.02 | 8.217 | 8.237 | 8.40 | 8.380 | 0.351 | 0.424 |
| 7 | Y | HA | 4.37 | 4.619 | 4.569 | 4.60 | 4.550 | 0.189 | 0.263 |
|   |   | HN | 8.00 | 8.191 | 8.291 | 8.38 | 8.280 | 0.436 | 0.523 |
| 8 | T | HA | 3.96 | 4.322 | 4.262 | 4.41 | 4.350 | 0.101 | 0.280 |
|   |   | HN | 7.54 | 8.199 | 7.959 | 8.00 | 8.240 | 0.308 | 0.531 |

## Table SII

**Table SII.** Experimental and simulation-derived HA and HN secondary chemical shifts.

| Residue number | Residue | Atom | Δδ experimental | Δδ simulation |
|---|---|---|---|---|
| 1 | A | HA | -0.66 | -0.3016 |
| 2 | S | HA | -0.08 | -0.051 |
|   |   | HN | 0.32 | 0.0932 |
| 3 | T | HA | -0.1 | -0.0469 |
|   |   | HN | -0.1 | -0.082 |
| 4 | T | HA | -0.11 | -0.0873 |
|   |   | HN | -0.15 | -0.0565 |
| 5 | T | HA | -0.16 | -0.0648 |
|   |   | HN | -0.31 | -0.1332 |
| 6 | N | HA | -0.23 | -0.0049 |
|   |   | HN | -0.38 | -0.163 |
| 7 | Y | HA | -0.23 | -0.0304 |
|   |   | HN | -0.38 | -0.0884 |
| 8 | T | HA | -0.45 | -0.148 |
|   |   | HN | -0.46 | -0.0402 |

# Figure S1

**Figure S1.** Superposition and WebLogo diagrams for each of the eight clusters identified by the dPCA for residues 5-8. The second column represents the superposition of 500 structures that belong to each cluster derived from the dPCA. The color coding is cyan for C atoms, blue for N atoms and red for O atoms. The third column consists of WebLogo diagrams of the per residue STRIDE-derived secondary structure assignments.

| Cluster | Superimposed structures | Secondary structure logo |
|---|---|---|
| 1<br><br>**829017 structures** | 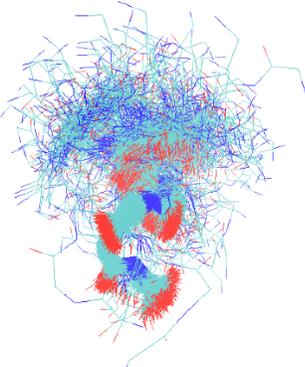 | 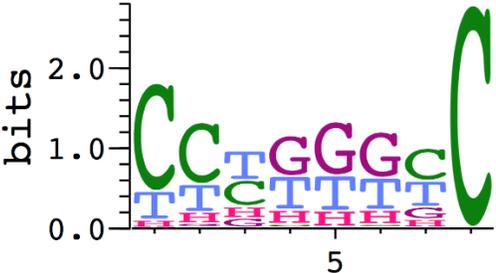 |
| 2<br><br>**627294 structures** | 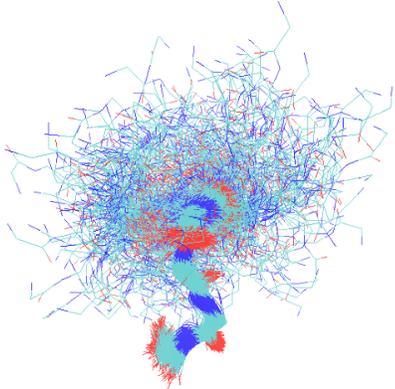 | 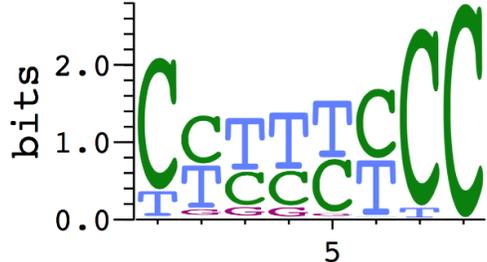 |
| **3**<br><br>**360755 structures** | 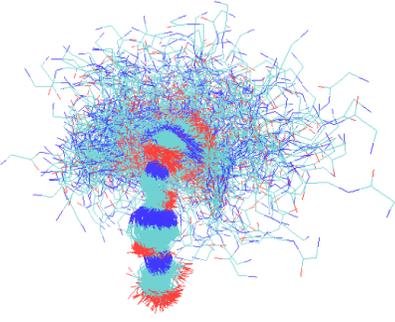 | 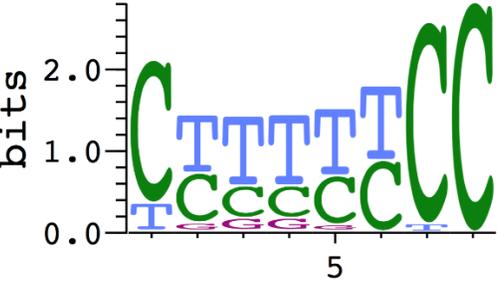 |
| **4**<br><br>**456049 structures** | 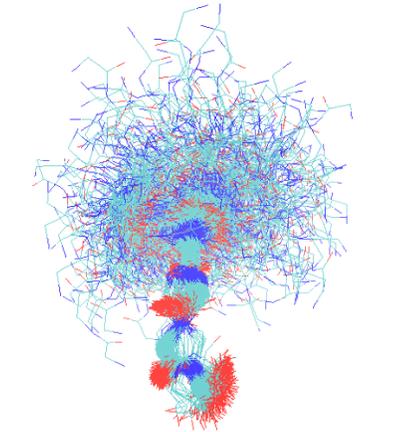 | 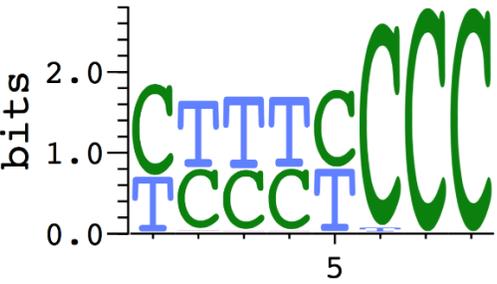 |
| **5**<br><br>**310782 structures** | 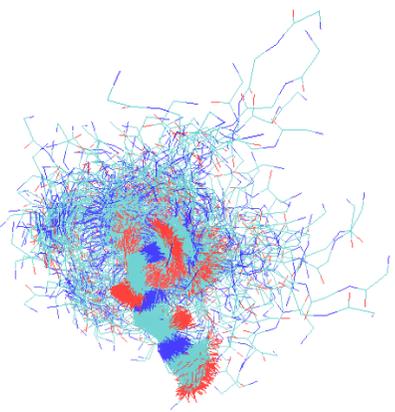 | 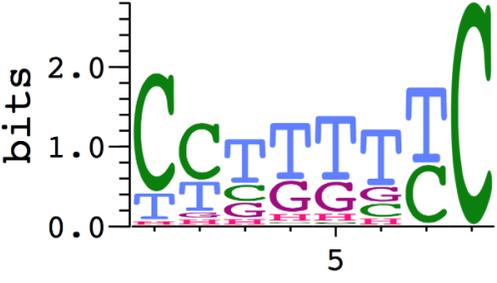 |

| 6

144079 structures | 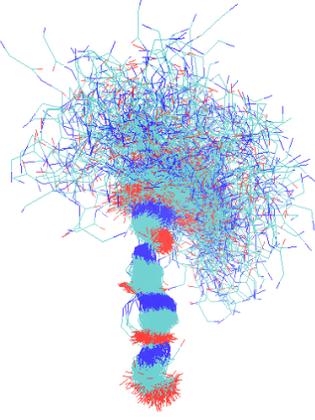 | 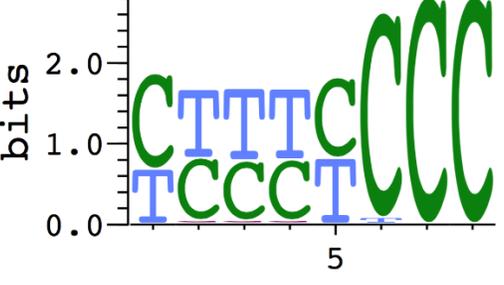 |
| --- | --- | --- |
| 7

128426 structures | 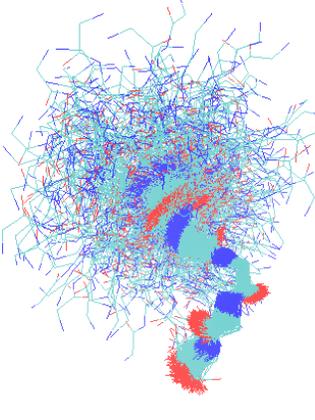 | 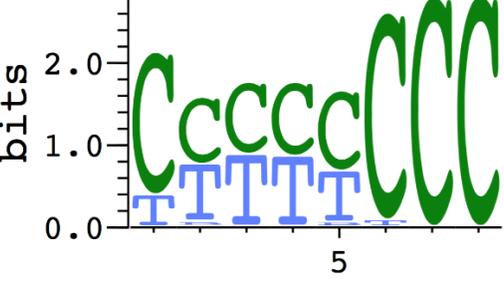 |
| 8

37878 structures | 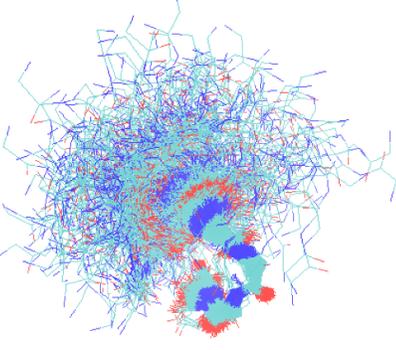 | 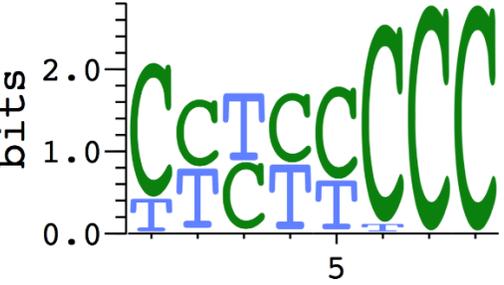 |